%% file: 0_main.tex
\documentclass[11pt]{article}

\usepackage[final]{acl}

\usepackage{times}
\usepackage{latexsym}

\usepackage[T1]{fontenc}

\usepackage[utf8]{inputenc}

\usepackage{microtype}

\usepackage{inconsolata}

\usepackage{graphicx}

\usepackage{amsmath}
\usepackage{amssymb}
\usepackage{booktabs}
\usepackage{multirow}
\usepackage{enumitem}
\usepackage{booktabs} 
\usepackage{tabularx} 
\usepackage{listings}
\title{RTLocating: Intent-aware RTL Localization for \\ Hardware Design Iteration}

\author{
 \textbf{Changwen Xing\textsuperscript{1,2}},
 \textbf{Yanfeng Lu\textsuperscript{2}},
 \textbf{Lei Qi\textsuperscript{2,3}},
 \textbf{Chenxu Niu\textsuperscript{4}},
\\
 \textbf{Jie Li\textsuperscript{4}},
 \textbf{Xi Wang\textsuperscript{1,2}\thanks{Corresponding author.}},
 \textbf{Yong Chen\textsuperscript{4}},
 \textbf{Jun Yang \textsuperscript{1,2}}
\\
\\
 \textsuperscript{1}School of Integrated Circuits, Southeast University, Nanjing, China,
 \\
 \textsuperscript{2}National Center of Technology Innovation for EDA, Nanjing, China,
 \\
 \textsuperscript{3}School of Computer Science and Engineering, Southeast University, Nanjing, China,
 \\
 \textsuperscript{4}Department of Computer Science, Texas Tech University, Lubbock, USA
\\
}

\begin{document}
\maketitle

\begin{abstract}
Industrial chip development is inherently \textbf{iterative}, favoring localized, intent-driven updates over rewriting RTL from scratch.
Yet most LLM-Aided Hardware Design (LAD) work focuses on \textit{one-shot} synthesis, leaving this workflow underexplored.
To bridge this gap, we \textbf{for the first time} formalize \(\Delta\)Spec-to-RTL localization, a multi-positive problem mapping natural language change requests ($\Delta$Spec) to the affected Register Transfer Level (RTL) syntactic blocks. 
We propose \textbf{RTLocating}, an intent-aware RTL localization framework, featuring a dynamic router that adaptively fuses complementary views from a textual semantic encoder, a local structural encoder, and a global interaction and dependency encoder (GLIDE).
To enable scalable supervision, we introduce \textbf{EvoRTL-Bench}, the first industrial-scale benchmark for intent-code alignment derived from OpenTitan's Git history, comprising \textbf{1,905} validated requests and \textbf{13,583} $\Delta$Spec-RTL block pairs. 
On EvoRTL-Bench, RTLocating achieves \textbf{0.568 MRR} and \textbf{15.08\% R@1}, outperforming the strongest baseline by \textbf{+22.9\%} and \textbf{+67.0\%}, respectively, establishing a new state-of-the-art for intent-driven localization in evolving hardware designs.
\end{abstract}

\section{Introduction}
\input{section/1_introduction}

\section{Related Work}
\input{section/2_related_works}

\section{Problem Formulation}
\label{sec:promblem_formulation}
\input{section/3_problem_formulation}

\section{Design \& Methodology}
\input{section/4_methodology}

\section{EvoRTL-Bench: Dataset Construction}
\label{sec:dataset}
\input{section/5_dataset}

\section{Evaluation}
\input{section/6_evaluation}

\section{Conclusion}
While industrial hardware development is inherently iterative, existing LLM-Aided Hardware Design (LAD) research predominantly targets one-shot synthesis. 
To bridge this gap, we formalize the \(\Delta\)Spec-to-RTL task and propose RTLocating, a framework that integrates textual semantics with local and design-wide structural signals through an intent-aware router. 
We further introduce EvoRTL-Bench, the first industrial-grade benchmark to enable fine-grained intent--RTL supervision. 
We hope this work encourages a paradigm shift in LAD: moving from static, one-shot generation toward evolution-aware, intent-guided iteration in real-world hardware design workflows.

\input{section/8_misc}

\bibliography{anthology}

\section*{Appendix}
\input{section/7_appendix}

\end{document}

%% file: section/1_introduction.tex
The escalating complexity of modern System-on-Chip (SoC) designs has substantially prolonged design and verification cycles, driving up iteration cost and time-to-market pressure~\cite{rhines2002dac}. As such, the semiconductor industry seeks to shorten the specification-to-silicon pathway. Recent progress in large language models (LLMs) for code understanding and generation~\cite{vaswani2017attention, koroteev2021bert} has catalyzed LLM-Aided Hardware Design (LAD), which aims to automate Register Transfer Level (RTL) development from natural-language specifications~\cite{9218553}.

However, most existing LAD research formulates hardware development as \textbf{one-shot synthesis}, where a model generates a self-contained, compilable RTL module from a natural-language specification (Spec), without leveraging existing design context or evolutionary history~\cite{thakur2024verigen, wang2024chatcpu}.
Constrained by the inherent ambiguity of natural language and the computational limits of standard attention over long contexts, prior approaches have primarily been validated on small academic benchmarks (e.g., RTLLM 2.0~\cite{liu2024openllm} $<200$ lines; VerilogEval~\cite{liu2023verilogeval} $<100$ lines), and they struggle to scale to real industrial settings (e.g., in OpenTitan~\cite{Opentitan2025}, where 83.7\% of Intellectual Property cores (IP) $>1000$ lines) with long-range dependencies and hierarchical organization. We further observe a fundamental mismatch between the prevailing one-shot abstraction in LAD and the industrial paradigm of \textbf{incremental iteration}: engineers rarely rewrite an entire design from scratch; instead, they apply minimal-perturbation updates to an existing RTL codebase in response to localized change requests (e.g., feature extensions, behavioral fixes, or micro-architectural optimizations), while keeping the rest of the system stable and verifiable~\cite{keating2002reuse}. 

\begin{figure}[t]
  \includegraphics[width=\columnwidth]{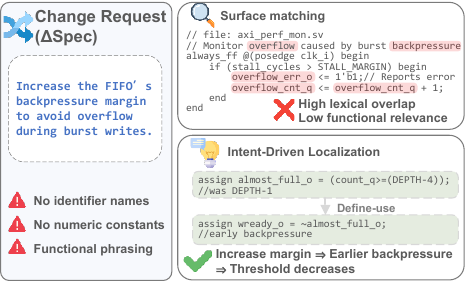}
  \caption{Intent-Driven RTL Change Localization vs. Surface Matching}
  \label{fig:surface_intent_compare}
\end{figure}
Consequently, for LAD to be truly practical in industry, its focus should shift from merely ``generating longer code'' to \textbf{intent-driven change reasoning}---deciding both \emph{where} and \emph{how} to change under explicit intent constraints.
Crucially, in large codebases, accurately identifying the impacted RTL scope (i.e., functional localization) is a critical prerequisite for producing reliable patches. Yet enabling this capability remains challenging due to two core barriers:
\textbf{1) Missing intent-to-RTL alignment supervision.} 
Industrial iteration demands mapping functional change requests ($\Delta$Spec) to the set of impacted RTL implementation locations (i.e., the modification scope). However, existing datasets lack explicit supervision that aligns high-level design intent with fine-grained RTL code snippets, preventing models from learning \emph{where} to localize modifications.
\textbf{2) Deficient intent-aware representation.} 
Effective localization requires positioning a block's functional role within the global design relative to the specific intent. Most existing approaches are confined to local syntax or surface semantics, and fail to support intent-conditioned cross-level reasoning. A lexical-functional mismatch example is illustrated in Figure~\ref{fig:surface_intent_compare}.

Motivated by these challenges, to our knowledge, we are the first to reformulate the hardware design iteration as an intent-driven functional localization task: given a functional change description ($\Delta$Spec), the goal is to identify and prioritize all RTL scope relevant to the change. To achieve this, we introduce \textbf{RTLocating}, the first intent-aware Spec-to-RTL localization framework that unifies textual semantics, local micro-architectural structure, and design-wide system topology. By employing a query-conditioned gating mechanism, RTLocating adaptively weights these complementary views, enabling accurate localization under industrial-scale design evolution.
Our contributions are summarized as follows:

\begin{enumerate}[leftmargin=*, nosep]

\item We develop a \textbf{function-aware hierarchical RTL representation} that integrates textual semantics, DFG-based local behavior,
and design-wide topological context, yielding unified embeddings for accurate intent-to-RTL alignment.

\item We construct \textbf{EvoRTL-Bench}, the first industrial-scale benchmark for intent-code alignment by mining the Git revision history of OpenTitan. It bridges the long-standing gap in function-level supervision and moves LAD research from static snapshots toward dynamic, evolution-based incremental iterations.

\item We design a \textbf{Design-Wide Topology Graph (DTG)} to encode full-design dependencies. By situating isolated RTL units within a global context, DTG enables long-range reasoning for cross-module modifications.

\item We introduce an \textbf{intent-aware adaptive router} that dynamically reweights the textual, local, and global encoders per change request, prioritizing the most informative features for robust localization.

\end{enumerate}

%% file: section/2_related_works.tex
\paragraph{LLM-Aided Hardware Design (LAD).} LAD aims to translate natural-language specifications into RTL implementations. Most prior work concentrates on \emph{from-scratch} RTL synthesis~\cite{pei2024betterv, liu2023chipnemo}, where LLMs are prompted or fine-tuned to generate self-contained modules and are evaluated primarily on small academic benchmarks~\cite{li2025deepcircuitx,  zou2024vgbench}. Complementary efforts study interactive and verification-oriented assistance~\cite{blocklove2023chip, yao2025hdldebugger, hu2024uvllm}. However, these settings diverge from industrial practice, where development relies on evolutionary, cross-module updates~\cite{saleh2006system}. This unexplored paradigm motivates our focus on intent-driven functional localization.

\paragraph{Natural Language to Code Localization.}
NL-to-code localization is well-established in software engineering, utilizing dense bi-encoders, such as CodeBERT~\cite{feng2020codebert}, GraphCodeBERT~\cite{guo2020graphcodebert} and UniXcoder~\cite{guo2022unixcoder} to map queries to code snippets. However, transferring these paradigms to industrial RTL iteration proves non-trivial: (i) existing models target sequential languages, failing to capture RTL-specific concurrency and timing semantics; (ii) benchmarks typically assume query-to-single-snippet matching, whereas functional changes inherently impact multiple distributed scopes; and (iii) evaluations focus on static snapshots, neglecting the evolutionary context of versioned histories. We address these gaps by formulating the $\Delta$Spec-to-RTL multi-positive ranking task and mining functional supervision from Git evolution.

\paragraph{Graph-based Code Representation.}
While the ``code-as-graph'' paradigm enhances code understanding via structural views (e.g., abstract syntax tree (AST), data flow graph (DFG)), its fine-grained nature (at the signal or operator level) poses challenges when scaled to industrial whole-design RTL~\cite{allamanis2017learning, yamaguchi2014modeling, zhou2019devign}. Specifically, the resulting graph explosion induces over-smoothing and noise accumulation in GNNs, hindering the capture of global functional semantics~\cite{li2018deeper, alon2020bottleneck, chen2020measuring}. To address this, we propose the Design-Wide Topology Graph (DTG). By adopting RTL syntactic blocks as primitive units, this structure incorporates global architectural context by elevating the level of abstraction, effectively enabling system-level functional representation.

\begin{figure*}[t]
    \centering
    \includegraphics[width=0.98\textwidth]{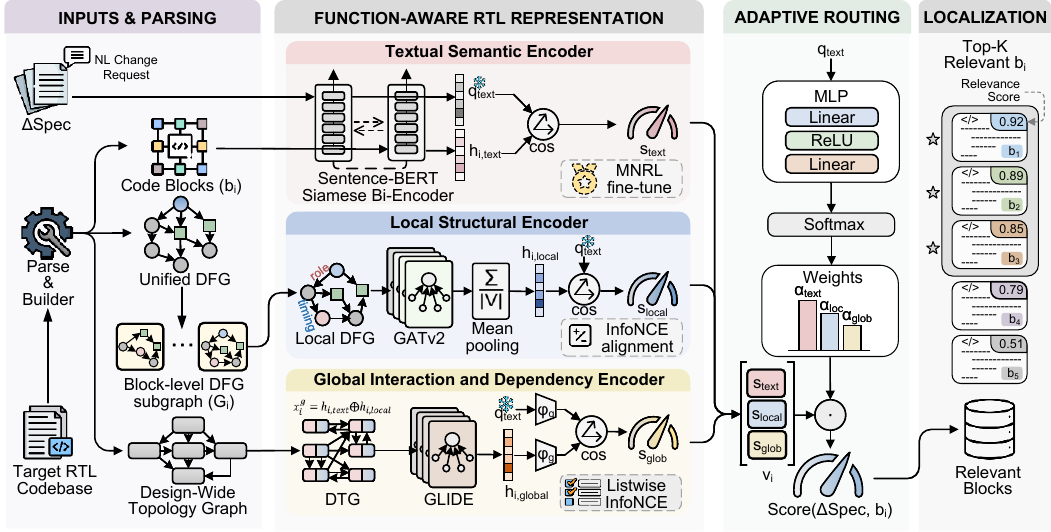}
    \caption{Overview of RTLocating Framework.}
    \label{fig:framework}
\end{figure*}

%% file: section/3_problem_formulation.tex

\paragraph{Task Definition.}
We formalize iterative RTL development as an intent-driven RTL block localization task.
Let $\mathcal{B}=\{b_1,b_2,\dots,b_N\}$ denote an RTL codebase snapshot. We define each $b_i \in \mathcal{B}$ as an \textbf{atomic syntactic block}, including \texttt{assign} statements and \texttt{always\_ff}/\texttt{always\_comb} procedures.
Given a natural-language change request $q$ (i.e., $\Delta\text{Spec}$), the goal is to identify the subset of affected blocks $\mathcal{G}\subseteq\mathcal{B}$.
This task is inherently multi-positive: a single intent often necessitates coordinated edits across distributed scopes.
Accordingly, the model should learn a ranking function where $\forall b^+ \in \mathcal{G}, \forall b^- \in \mathcal{B}\setminus\mathcal{G}: \text{rank}(b^+) < \text{rank}(b^-)$.

\paragraph{Scoring and Optimization.}
We learn a relevance scoring function $s_\theta(q,b)$ between a change request $q$ and an RTL block $b$.
To encourage positives $\mathcal{G}_q$ to be ranked above negatives $\mathcal{N}_q$, we minimize the averaged pairwise margin ranking loss:
\begin{equation}
\begin{split}
\mathcal{L}_{\text{rank}}(q)
= & \frac{1}{|\mathcal{G}_q|\,|\mathcal{N}'_q|}
\sum_{b_p\in\mathcal{G}_q}
\sum_{b_n\in\mathcal{N}'_q} \\
& \max\bigl(0,\ \gamma - s_\theta(q,b_p) + s_\theta(q,b_n)\bigr),
\end{split}
\end{equation}
where $\gamma$ is a margin hyperparameter and $\mathcal{N}'_q \subset \mathcal{N}_q$ denotes sampled negatives.

%% file: section/4_methodology.tex
We propose \textbf{RTLocating}, an intent-driven functional localization framework for industrial RTL design iteration, as shown in  Figure~\ref{fig:framework}.
Given a natural-language change request ($\Delta\text{Spec}$) and a target codebase, RTLocating decomposes the design into block units, deriving intra-block data-flow graphs (DFG) and a global design-wide topology graph (DTG). 
We construct multi-dimensional representations via three complementary encoders to align textual semantics, encode local structural dependencies, and model system-level interactions (via GLIDE). Each encoder is trained with a dedicated supervision objective reflecting its localization role.
Finally, an \textbf{intent-aware} router adaptively weights these loss signals to retrieve the top-ranked blocks for modification.

\subsection{Multi-Dimensional RTL Representation}
\label{sec:Multi-View RTL Representation}

\subsubsection{Textual Semantic Encoder}
\label{sec:text_expert}
The Textual Semantic Encoder aligns change requests $q$ (i.e., $\Delta\text{Spec}$) and RTL blocks $b_i$ in a shared semantic space via a parameter-shared bi-encoder (initialized with SBERT~\cite{reimers2019sentence}).
This yields embeddings $\mathbf{q}_{\text{text}},\,\mathbf{h}_{i,\text{text}}\in\mathbb{R}^{d}$, supporting efficient offline indexing.

\paragraph{Fine-tuning Strategy.} To capture RTL vernacular, we fine-tune the Textual Semantic Encoder using Multiple Negatives Ranking Loss (MNRL). 
Given a batch of $k$ pairs $\{(q_j,b_j^{+})\}_{j=1}^{k}$, we minimize:
\begin{equation}
\mathcal{L}_{\text{text}}
=
-\frac{1}{k}\sum_{j=1}^{k} \log
\frac{\exp(\mathrm{sim}(q_j,b_j^{+})/\tau)}
{\sum_{l=1}^{k}\exp(\mathrm{sim}(q_j,b_l^{+})/\tau)}.
\end{equation}
where $\mathrm{sim}(\cdot, \cdot)$ denotes cosine similarity and $\tau$ is the temperature hyperparameter.
Since MNRL treats in-batch samples as negatives, it risks penalizing valid positives in multi-positive settings. We mitigate this by ensuring each batch contains at most one positive block per $\Delta\text{Spec}$.

\subsubsection{Local Structural Encoder}
\label{sec:local_encoder}

Direct encoding of the full IP-level DFG is computationally prohibitive due to its fine granularity. Consequently, we adopt a ``divide-and-conquer'' strategy, decomposing the IP into a collection of subgraphs by constructing a block-local DFG $\{G_i=(V_i, E_i)\}$ for each atomic syntactic RTL block. This approach effectively captures intra-block logical structures while circumventing computational bottlenecks.

\paragraph{Node and Edge Initialization.}
\noindent\textbf{Nodes.} $V_i$ consists of \emph{signals} and \emph{operators}.
For each node $v$, we construct the input feature $\mathbf{x}_v$ by concatenating the embeddings of:
(i) a hashed identifier,
(ii) a node category $\mathrm{cat}(v)\in\mathcal{C}$ (e.g., parameter, port, operators),
(iii) an operator type (e.g., MemberAccess, UnaryOp, BinaryOp), 
(iv) a normalized bit-width scalar $\widehat{w}(v)$.
Missing attributes are mapped to a learnable default embedding.

\noindent\textbf{Edges.} Each edge $e=(u \to v)$ carries a dependency role $r(e)$ (e.g., clock, event, predicate) and a timing class $t(e)$ (e.g., combinational vs. sequential).
These discrete attributes are embedded and concatenated to form the edge feature $\mathbf{x}_e$.

\paragraph{Graph Encoding.}
To model heterogeneous intra-block dependency patterns, we employ GATv2~\cite{brody2021attentive} to encode $G_i$, leveraging its ability to modulate attention scores via edge attributes.
After $L$ layers of message passing, we obtain the contextualized node embeddings $\{\mathbf{h}_v^{(L)}\}_{v \in V_i}$ and derive the block representation via global mean pooling followed by a projection head:
\begin{equation}
    \mathbf{h}_{i,\text{local}} = \mathrm{Normalize}\left( \mathbf{W}_p \left( \frac{1}{|V_i|} \sum_{v \in V_i} \mathbf{h}_v^{(L)} \right) \right),
\end{equation}
where $\text{W}_p$ projects the pooled graph embedding into the shared embedding space, and $\text{Normalize}(\cdot)$ denotes $\ell_2$-normalization. With the Textual Semantic Encoder frozen, we optimize the Local Structural
Encoder using an InfoNCE objective to align $\mathbf{h}_{i,\text{local}}$ with its corresponding textual intent
$\mathbf{q}_{\text{text}}$.

\subsubsection{Global Interaction and Dependency Encoder}
\label{sec:global_expert}
While the local encoder captures intra-block features, it overlooks the holistic system architecture. Therefore, we propose the \textbf{GL}obal \textbf{I}nteraction and \textbf{D}ependency \textbf{E}ncoder (GLIDE), a graph-based encoder that operates on the Design-Wide Topology Graph (DTG) to capture system-level interactions among RTL blocks. By embedding each block within this global interaction context, GLIDE characterizes its functional role beyond local data-flow dependencies. 

\paragraph{Design-Wide Topology Graph (DTG) Construction.} 
DTG is constructed for each design version, denoted as $G^{\text{glob}}=(V^{g},E^{g})$. Each node $v_i \in V^{g}$ corresponds to an atomic syntactic block. 
Node features $\mathbf{x}^{g}_{i}$ are initialized by concatenating the textual and local structural embeddings, providing semantic and behavioral priors for subsequent global message passing:
\begin{equation}
\mathbf{x}^{g}_{i} = \mathbf{h}_{i,\text{text}} \,\Vert\, \mathbf{h}_{i,\text{local}} .
\end{equation}
Edges encode directed define--use dependencies ($u \!\rightarrow\! v$ if $b_u$ drives $b_v$). To accommodate multiple distinct inter-block dependencies, $G^{\text{glob}}$ is formulated as a typed directed multigraph; parallel dependencies are preserved as distinct edges annotated with dependency roles and timing classes.

\paragraph{Context Propagation and Alignment.} 
We encode $G^{\text{glob}}$ using GLIDE, which is implemented as a query-agnostic multi-layer GATv2-based network~\cite{brody2021attentive} and stabilized by residual connections and layer normalization. 
It incorporates embedded edge attributes as edge features in attention, producing design-wide contextualized embeddings $\{\mathbf{h}_{i,\text{global}}\}_{i \in \mathcal{V}}$
 
We align the $\mathbf{h}_{i,\text{global}}$ with textual queries $\mathbf{q}_{\text{text}}$ in a shared metric space using projection heads $\phi_g(\cdot)$ and $\phi_q(\cdot)$, while keeping the Textual Semantic Encoder frozen. 
To ensure fine-grained discrimination, we construct a candidate set $\mathcal{C}_p=\{p\}\cup \mathcal{N}_p$ for each target block $p \in \mathcal{P}$, where negatives $\mathcal{N}_p$ are sampled from unaffected blocks within the same design. 
We optimize the model via the following InfoNCE objective, using temperature-scaled cosine similarity $\ell(q_\text{text}, \cdot)$ as the alignment score:
\begin{equation}
\mathcal{L}_{\text{global}}
=
-\frac{1}{|\mathcal{P}|}\sum_{p\in\mathcal{P}}
\log
\frac{\exp(\ell(q_\text{text}, p))}
{\sum_{c\in\mathcal{C}_p}\exp(\ell(q_\text{text}, c))} .
\end{equation}

\subsection{Intent-Aware Adaptive Routing}
Static fusion strategies (e.g., fixed weighted averaging) fail to adapt to varying retrieval intents, which range from specific signal naming (\textit{text-dominant}) to complex logic changes (\textit{structure-dominant}). To address this, we propose an adaptive router that dynamically synthesizes heterogeneous signals from different experts.

First, we condense the evidence for each pair $(\mathbf{q}_{\text{text}}, b_i)$ into a vector $\mathbf{v}_i \in \mathbb{R}^{3}$. 
This vector aggregates normalized similarities from the text, local, and global views:
\begin{equation}
    \mathbf{v}_i = \bigl[ s_{\text{txt}},\, s_{\text{loc}},\, s_{\text{glob}} \bigr]^\top,
\end{equation}
where $s_{\text{txt}}$, $s_{\text{loc}}$, and $s_{\text{glob}}$ denote cosine scores derived from their respective encoders. 
Next, instead of using fixed weights, we employ a lightweight router to predict importance weights $\boldsymbol{\alpha}$ directly from the query embedding $\mathbf{q}_{\text{text}}$:
\begin{equation}
\boldsymbol{\alpha}=\mathrm{Softmax}\!\left(\mathrm{MLP}_{\text{route}}(\mathbf{q}_{\text{text}})\right)\in\mathbb{R}^{3}.
\end{equation}
The final relevance is scored via adaptive aggregation: $\mathrm{Score}(\mathbf{q}_{\text{text}}, b_i) = \boldsymbol{\alpha}^\top \mathbf{v}_i$.
This design decouples weight generation from candidate interaction; $\boldsymbol{\alpha}$ is computed once per query, ensuring efficient ranking over large candidate pools.

%% file: section/5_dataset.tex
\begin{table}[t]
\centering
\small
\setlength{\tabcolsep}{4pt}
\begin{tabularx}{\columnwidth}{@{}lcc>{\raggedleft\arraybackslash}X@{}}
\toprule
\textbf{Tier} & \textbf{LOC Range} & \textbf{\#IPs} & \textbf{Share} \\ \midrule
Lightweight     & $<1$k    & 6  & 16.2\% \\
Small/Utility   & 1k--5k   & 15 & 40.6\% \\
Medium-Scale    & 5k--10k  & 8  & 21.6\% \\
Large-Scale     & 10k--20k & 6  & 16.2\% \\
High-Complexity & $>20$k   & 2  & 5.4\%  \\ \midrule
\textbf{Total}  & --       & \textbf{37} & \textbf{100.0\%} \\
\bottomrule
\end{tabularx}
\caption{OpenTitan IP Complexity Statistics (LOC).}
\label{tab:opentitan_complexity_dist}
\end{table}


To benchmark intent-driven functional localization under realistic RTL evolution, we introduce \textbf{EvoRTL-Bench},
mined from the Git history of OpenTitan~\cite{Opentitan2025}. OpenTitan is an open-source silicon Root of Trust (RoT)
initiative with production-scale IPs and standardized development practices. As shown in Table~\ref{tab:opentitan_complexity_dist},
its IPs span a wide range of code sizes measured in lines of code (LOC), and its functional diversity statistics are reported in Appendix~\ref{app:OpenTitan}. 
EvoRTL-Bench leverages this high-quality source to provide multi-positive supervision, pairing natural-language change requests with ground-truth affected RTL blocks. Our construction pipeline comprises two phases.

\paragraph{Phase 1: Revision Mining and Intent Extraction.} 
We traverse the Git history to identify commits modifying SystemVerilog design files, explicitly excluding testbenches and auto-generated artifacts. 
Leveraging a large language model (LLM), we transform raw commit messages into structured $\Delta$Specs comprising: (i) behavioral summaries ($S_{\text{old}}, S_{\text{new}}$), (ii) functional context, and (iii) explicit change intent. 
To ensure precise intent-code alignment, we strictly filter for functional alterations, pruning non-functional maintenance (e.g., refactoring, formatting) and ambiguous entries. 
Detailed prompts and quality control protocols are provided in Appendix \ref{app:EvoRTL}.

\paragraph{Phase 2: Syntax-Aware Block Localization.} 
To establish the ground-truth modification scope, we reconstruct pre- and post-revision snapshots to compute line-level diff hunks. 
These textual changes are then mapped to their enclosing atomic syntactic blocks within the AST (as defined in \S~\ref{sec:promblem_formulation}).

\paragraph{Dataset Summary.} 
The curated dataset comprises 1,905 validated $(\Delta\text{Spec}, \text{block-set})$ instances. 
Flattening these sets yields a final corpus of 13,583 $(\Delta\text{Spec}, \text{block})$ pairs.

%% file: section/6_evaluation.tex
\begin{table*}[t]
\centering
\renewcommand{\arraystretch}{1.1} 
\setlength{\tabcolsep}{2pt} 
\footnotesize 
\begin{tabular*}{\textwidth}{@{\extracolsep{\fill}}l|c|cc|ccc|ccc}
\toprule
\multirow{2}{*}{\textbf{Model}} & \multirow{2}{*}{\textbf{Setting}} & \multirow{2}{*}{\textbf{MRR}} & \multirow{2}{*}{\textbf{MAP}} & \multicolumn{3}{c|}{\textbf{Recall (\%)}} & \multicolumn{3}{c}{\textbf{Hit (\%)}} \\
 & & & & \textbf{R@1} & \textbf{R@5} & \textbf{R@10} & \textbf{H@1} & \textbf{H@5} & \textbf{H@10} \\
\midrule
\multicolumn{10}{l}{\textit{Lexical \& General Dense Retrievers}} \\
BM25 & - & 0.421 & 0.292 & 8.33 & 28.02 & 33.90 & 28.18 & 59.09 & 65.45 \\
all-MiniLM-L6-v2 & Off-the-shelf & 0.373 & 0.229 & 7.39 & 25.56 & 31.04 & 22.73 & 54.55 & 64.55 \\
\midrule
\multicolumn{10}{l}{\textit{Code-Specific Pre-trained Models}} \\
\multirow{2}{*}{CodeBERT} & Off-the-shelf & 0.110 & 0.046 & 0.86 & 2.21 & 4.98 & 5.45 & 15.45 & 21.82 \\
 & Fine-tuned & 0.204 & 0.100 & 3.08 & 9.20 & 11.48 & 11.82 & 26.36 & 37.27 \\
\addlinespace[1pt]
\multirow{2}{*}{GraphCodeBERT} & Off-the-shelf & 0.229 & 0.115 & 5.52 & 10.35 & 15.55 & 13.64 & 33.64 & 43.64 \\
 & Fine-tuned & 0.416 & 0.262 & 7.71 & \underline{30.37} & 34.00 & 26.36 & \underline{66.36} & \underline{73.64} \\
\addlinespace[1pt]
\multirow{2}{*}{UniXcoder} & Off-the-shelf & 0.334 & 0.173 & 5.74 & 16.66 & 24.08 & 24.55 & 55.45 & 64.55 \\
 & Fine-tuned & 0.427 & 0.250 & 8.01 & 26.49 & 34.67 & \underline{32.73} & 64.55 & \underline{73.64} \\
\midrule
\multicolumn{10}{l}{\textit{Sparse--Dense Hybrid Retriever}} \\
BM25 + UniX (FT) & - & \underline{0.462} & \underline{0.324} & \underline{9.03} & 29.10 & \underline{35.45} & 31.82 & 60.91 & 72.73 \\
\midrule
\textbf{\textbf{RTLocating} (Ours)} & - & \textbf{0.568} & \textbf{0.360} & \textbf{15.08} & \textbf{32.42} & \textbf{36.34} & \textbf{48.18} & \textbf{70.00} & \textbf{73.64} \\
\emph{vs. Best Baseline} & & \emph{+22.9\%} & \emph{+11.1\%} & \emph{+67.0\%} & \emph{+6.7\%} & \emph{+2.5\%} & \emph{+47.2\%} & \emph{+5.5\%} & \emph{+0.0\%} \\
\bottomrule
\end{tabular*}
\caption{Main Retrieval Performance of RTLocating and Baselines.}
\label{tab:main_results}
\end{table*}

\subsection{Experimental Setup}

\paragraph{Dataset and Splitting.}
We evaluate RTLocating on  EvoRTL-Bench (\S\ref{sec:dataset}) using an IP-disjoint split to test cross-design generalization: models are trained on a subset of IPs and evaluated on held-out IPs, preventing reliance on IP-specific identifiers or artifacts. The dataset comprises 13,583 $(\Delta\text{Spec}, \text{block})$ pairs, partitioned into 9,700/2,035/1,848 for train/validation/test.

\paragraph{Baselines.} We benchmark RTLocating against four categories:
\begin{itemize}[leftmargin=*, noitemsep, topsep=0pt, parsep=0pt]
    \item \textbf{Lexical IR.} \textbf{BM25}~\cite{robertson2009probabilistic}: Ranks blocks via sparse lexical overlap.
    \item \textbf{General Dense.} \textbf{all-MiniLM-L6-v2}~\cite{allminilmL6v2_modelcard}: A domain-agnostic baseline to gauge task difficulty.
    \item \textbf{Code-Pretrained.} \textbf{CodeBERT}~\cite{feng2020codebert}, \textbf{GraphCodeBERT}~\cite{guo2020graphcodebert} (data-flow aware), and \textbf{UniXcoder}~\cite{guo2022unixcoder} (unifying AST modalities) evaluated in both \textbf{zero-shot} and \textbf{fine-tuned} modes.
    \item \textbf{Sparse--Dense Hybrid.} \textbf{BM25 + UniXcoder (Fine-tuned)}: A practical industry-style pipeline that fuses sparse and dense scores to combine lexical and semantic matching.
\end{itemize}
All neural baselines operate as bi-encoders. Their fine-tuned variants share the identical objective, hyperparameters, and cosine ranking protocol as the Textual Semantic Encoder.


\paragraph{Implementation Details.}
Implemented on NVIDIA H20 GPUs, the Textual Semantic Encoder fine-tunes \texttt{all-MiniLM-L6-v2} (max len $384$) for 5 epochs. The Local Structural Encoder and GLIDE employ GATv2-based encoders, optimized via standard InfoNCE and listwise InfoNCE objectives, respectively. 
Finally, the Intent-Aware Dynamic Router utilizes a lightweight MLP to predict expert weights, trained using margin ranking loss.
For the construction of the  EvoRTL-Bench dataset, we utilized \texttt{ChatGPT-4.1}.
Comprehensive hyperparameters are detailed in Appendix~\ref{sec:imp_details}.

\paragraph{Evaluation Metrics.} 
We evaluate localization performance using four metrics: (i) \textbf{MRR}, the average reciprocal rank of the first relevant block; (ii) \textbf{MAP}, the mean average precision across all ground-truth blocks; (iii) \textbf{Recall@k}, the proportion of ground-truth blocks recovered in the top-$k$ candidates; and (iv) \textbf{Hit@k}, the percentage of queries where at least one relevant block appears in the top-$k$. We use $k \in \{1, 5, 10\}$ and report results averaged over three random seeds.

\subsection{Main Retrieval Performance}
\label{sec:main_results}
\paragraph{Main Results.} As shown in Table~\ref{tab:main_results}, RTLocating establishes a new state-of-the-art, outperforming the strongest baseline (Sparse--Dense Hybrid Retriever) by a substantial margin of +0.106 points in MRR (\textbf{+22.9\%}), and +0.036 points in MAP (\textbf{+11.1\%}), which demonstrates superior global ranking quality. Most notably, on the strictest top-ranked metrics, RTLocating delivers remarkable relative gains over the best baseline: \textbf{+67.0\%} in Recall@1 and \textbf{+47.2\%} in Hit@1.
As $k$ increases, performance differences shrink and Hit@10 becomes comparable across the strongest methods. The main improvement of RTLocating is concentrated at small cutoffs, where it more reliably promotes relevant blocks to the highest ranks.

\paragraph{Analysis.}
The results yield three critical insights into hardware design location:

\noindent\textbf{1) The Software-Hardware Domain Gap.}
Off-the-shelf code retrievers transfer poorly to RTL (e.g., CodeBERT: 0.110 MRR), confirming that pre-training fails to capture RTL semantics without explicit domain adaptation.

\noindent\textbf{2) Fine-tuning hits a ``Lexical Ceiling''.}
While fine-tuning boosts dense retrievers to BM25 performance ($\sim$0.42 MRR), lexical baselines remain highly competitive on strict metrics (MAP/R@1). This implies standard global alignment plateaus at keyword matching, failing to distinguish fine-grained functional nuances.

\noindent\textbf{3) Fine-grained Structure vs. Design-wide Context.}
Zero-shot GraphCodeBERT outperforms CodeBERT (0.229 vs. 0.110 MRR), demonstrating the benefit of data-flow inductive bias. However, the significant gap to RTLocating (0.568 MRR) reveals that token-level modeling is insufficient; capturing high-level, design-wide topology is essential for accurate localization.

\subsection{Efficiency Analysis}
\begin{table}[htbp]
    \centering
    \small 
    \begin{tabular}{l c c}
        \toprule
        \textbf{Model} & \textbf{Params (M)} & \textbf{Latency (ms)} \\
        \midrule
        CodeBERT & 124.65 & 7.23 \\
        GraphCodeBERT & 124.65 & 7.14 \\
        UniXcoder & 125.93 & 7.15 \\
        Sparse--Dense Hybrid & 125.93 & 14.32 \\
        \textbf{RTLocating (Ours)} & \textbf{23.04} & \textbf{6.25} \\
        \bottomrule
    \end{tabular}
    \caption{Comparison of Model Parameters and Inference Latency.}
    \label{tab:performance_comparison}
\end{table}

Table~\ref{tab:performance_comparison} summarizes model complexity and per-query inference latency. 
Remarkably, RTLocating achieves an \textbf{81.5\%} parameter reduction ($5.4\times$ compression) compared to transformer-based baselines (23.04 M vs. $\sim$125 M),
while achieving stronger retrieval performance (Table~\ref{tab:main_results}).
This compact design also yields the lowest latency (6.25\,ms), whereas the sparse---dense hybrid retriever incurs substantially higher cost (14.32\,ms) due to its dual-pipeline overhead. 
Overall, RTLocating achieves a favorable accuracy--efficiency balance for large-scale, resource-constrained industrial localization.

\subsection{Effect of Representations and Routing Mechanism}
\label{sec:ablation}
\begin{figure}[t]
  \includegraphics[width=\columnwidth]{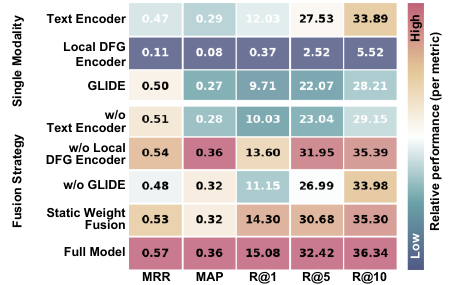}
  \caption{RTLocating Ablation Heatmap.}
  \label{fig:ablation}
\end{figure}

We quantify the contribution of each representation view and the intent-aware router via ablations, as summarized in Figure~\ref{fig:ablation}.

\paragraph{Design-wide topology is a strong standalone signal.}
The GLIDE-only variant outperforms the Text-only variant in MRR (0.501 vs.\ 0.469), indicating that design-wide dependencies provide informative system context. Removing GLIDE causes the largest drop (MRR 0.568$\rightarrow$0.478; MAP 0.360$\rightarrow$0.316), highlighting the importance of global topology for localization.

\paragraph{Text provides a precise semantic anchor.}
Despite higher overall ranking quality, GLIDE is weaker on top-1 identification than text (R@1 9.71 vs.\ 12.03), suggesting that textual cues are particularly useful for pinpointing the exact edit target among closely related units.


\paragraph{Local structure mainly helps refine the top ranks.}
Local structure alone is insufficient (MRR 0.112), but adding it to text+global yields a consistent gain at R@1 (13.60$\rightarrow$15.08), indicating its role as a refinement signal once relevant candidates are retrieved.


\paragraph{Dynamic routing is necessary.}
Static fusion underperforms dynamic routing (MRR 0.53 vs.\ 0.57) and even trails the w/o Local variant (MRR 0.54), implying that local structural cues are not uniformly beneficial. The router adaptively reweights encoders per query, mitigating noise from ubiquitous micro-structures.

\subsection{Interpretability of Intent-Aware Routing}
\label{sec:moe_analysis}
\begin{figure}[t]
  \includegraphics[width=\columnwidth]{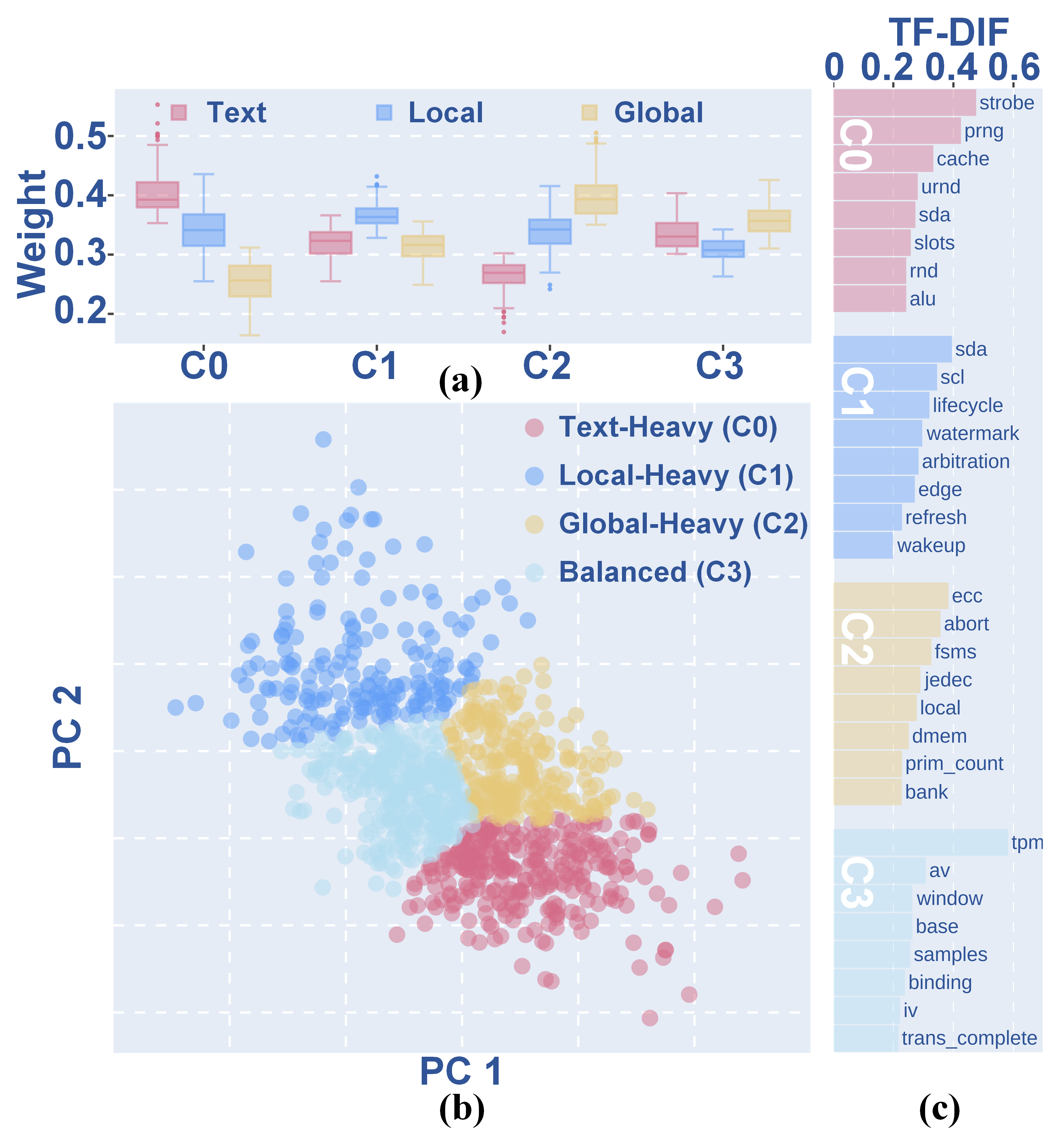}
    \caption{Intent-aware Routing Analysis: (a) encoder weight distributions; (b) PCA of weight clusters; (c) top TF-IDF keywords.}
  \label{fig:expert_clusters}
\end{figure}

We analyze routing by clustering query-conditioned expert weights with K-Means.
We set $K{=}4$ as the smallest value that yields interpretable modes---\emph{text-dominant}, \emph{local-dominant},
\emph{global-dominant}, and \emph{balanced}. Figure~\ref{fig:expert_clusters} shows clear separation in encoder preferences (boxplots).
To characterize each mode, we extract cluster-specific TF-IDF keywords after filtering generic HDL syntax
(e.g., \textit{module}, \textit{signal}) and ubiquitous hardware terms (e.g., \textit{bus}, \textit{register}).

\paragraph{C0: Terminology-driven (Text-heavy).}
C0 assigns the largest weight to the Textual Semantic Encoder (median $\alpha_{\text{text}}\!\approx\!0.40$).
Its queries contain distinctive domain terms and acronyms (e.g., \textit{strobe}, \textit{prng/urnd/rnd}, \textit{subbytes})
that map to well-scoped functionalities within the RTL corpus.
Here, lexical/semantic cues are often sufficient to disambiguate relevant units, and the router correspondingly down-weights
structural views.


\paragraph{C1: Interaction-driven (Local-heavy).}
C1 assigns the highest weight to the Local Structural Encoder (median $\alpha_{\text{loc}}\!\approx\!0.37$).
Queries in this cluster focus on protocol- and timing-sensitive behaviors (e.g., \textit{sda}, \textit{scl}, \textit{arbitration}, \textit{watermark}),
which often hinge on fine-grained intra-block signal interactions. Accordingly, the router places greater emphasis on the local view for these intents.

\paragraph{C2: Architecture-driven (Global-heavy).}
C2 favors the GLIDE (median $\alpha_{\text{glob}}\!\approx\!0.40$) and is dominated by system-level control and memory-subsystem intents.
Keywords such as \textit{dmem}, \textit{bank}, \textit{jedec}, \textit{ecc}, \textit{abort}, and \textit{fsm}
refer to architectural components and control paths whose relevance is largely determined by cross-block define--use relations rather than distinctive local syntax.
Accordingly, the router prioritizes GLIDE to leverage design-wide dependency context.

\subsection{Robustness to Identifier Anonymization}
\label{sec:mask_analysis}


To assess the extent to which RTLocating relies on structural semantics rather than lexical surface forms, we anonymize test-time RTL candidates by deterministically remapping all user-defined
identifiers (e.g., signals and instances) to neutral placeholders (e.g., \texttt{VAR\_i}), while preserving SystemVerilog keywords
and data-flow structure. This removes identifier-level lexical cues without retraining, leaving primarily structural/topological
information.

\begin{figure}[t]
    \centering
  \includegraphics[width=\columnwidth]{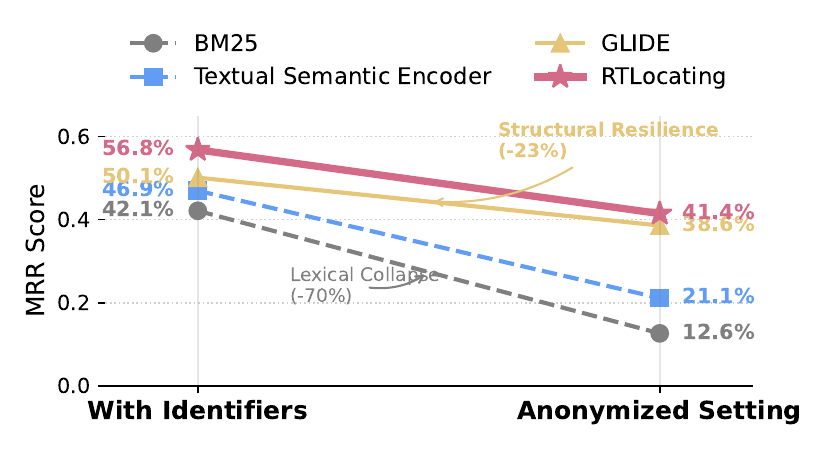}
  \caption{Robustness Analysis via Slope Chart.}
  \label{fig:robustness_slope}
\end{figure}

As shown in Figure~\ref{fig:robustness_slope}, lexical baselines (BM25 and the Textual Semantic Encoder) drop markedly under
anonymization, consistent with their reliance on surface overlap. GLIDE degrades less, indicating that design-wide dependencies
remain informative when identifier semantics are removed. Notably, RTLocating achieves the strongest masked-setting
performance, suggesting our model effectively captures intrinsic structural semantics rather than overfitting to variable naming conventions.


%% file: section/8_misc.tex
\section*{Limitations}
This work focuses on the \emph{localization} stage of iterative LAD, i.e., identifying the impacted RTL implementation scope given a change intent. While accurate localization is a necessary prerequisite for reliable updates, we have not built or evaluated a complete LLM-driven iterative design pipeline that performs subsequent edit generation, validation, and repair.
As a result, the end-to-end effectiveness of integrating RTLocating into a full specification-to-patch workflow remains to be established, and future work should assess how localization quality translates into downstream editing success under
realistic verification constraints.

\section*{Ethics Consideration}
\label{sec:ethics}

EvoRTL-Bench is mined from the publicly available OpenTitan repository history.
To mitigate privacy risks, we do not collect or release any author-identifying metadata (e.g., names, email addresses,
usernames/handles) from commits or review discussions.
For the natural-language fields (e.g., change requests), we apply automatic filters to remove email-like strings,
\texttt{@}-mentions, and signature lines (e.g., Signed-off-by, Reviewed-by) when present.
We do not observe offensive content in the curated corpus; nevertheless, we perform a lightweight keyword-based screening and
manually inspect flagged cases before release.
The benchmark is intended for research on intent-driven RTL localization and evolution-aware LAD.
It should not be used to facilitate unauthorized vulnerability discovery or exploit development.

%% file: section/7_appendix.tex
\appendix
\section{Detailed Implementation Settings}
\label{sec:imp_details}

This section details the experimental configurations to facilitate reproducibility. Our framework is implemented using PyTorch and PyTorch Geometric on NVIDIA H20 GPUs, with a fixed random seed of $36$ for all experiments. Table \ref{tab:hyperparameters} provides a comprehensive breakdown of the architectural hyperparameters, optimization settings, and training specifics for each component.

\begin{table}[h]
\centering
\small
\renewcommand{\arraystretch}{1.2}
\begin{tabularx}{\columnwidth}{@{}l X l@{}} 
\toprule
\textbf{Component} & \textbf{Hyperparameter} & \textbf{Value} \\ \midrule
\multicolumn{3}{@{}l@{}}{\textit{\textbf{General Settings}}} \\
& Frameworks & PyTorch, PyG \\
& Random Seed & 36 \\
& Optimizer & AdamW \\ \midrule

\multicolumn{3}{@{}l@{}}{\textit{\textbf{Text Expert}}} \\
& Base Model & \texttt{MiniLM-L6} \\
& Max Len & 384 \\
& Learning Rate & $2\mathrm{e}{-5}$ \\
& Batch Size & 32 \\
& Epochs & 5 \\ \midrule

\multicolumn{3}{@{}l@{}}{\textit{\textbf{Local Structural Expert}}} \\
& Architecture & 3-layer GATv2 \\
& Hidden Dim ($h$) & 128 \\
& Dropout & 0.3 \\
& Learning Rate & $5\mathrm{e}{-4}$ \\
& Warmup & 10\% (Linear) \\
& Loss & InfoNCE \\ \midrule

\multicolumn{3}{@{}l@{}}{\textit{\textbf{Global Expert (GLIDE)}}} \\
& Architecture & 2-layer GATv2 \\
& Node Hidden ($h$) & 256 \\
& Edge Dim & 32 \\
& Proj. Heads & $384\!\to\!128$ \\
& Learning Rate & $5\mathrm{e}{-4}$ \\
& Loss & Listwise InfoNCE \\ \midrule

\multicolumn{3}{@{}l@{}}{\textit{\textbf{Intent-Aware Router}}} \\
& Structure & MLP ($128\!\to\!3$) \\
& Dropout & 0.3 \\
& Loss & Margin ($\gamma=0.5$) \\ \bottomrule
\end{tabularx}
\caption{Detailed hyperparameters and configurations.}
\label{tab:hyperparameters}
\end{table}

\section{OpenTitan Dataset Characteristics}
\label{app:OpenTitan}

To quantify the engineering scale of our dataset and demonstrate its representativeness of industrial-grade IP complexity, we employ SystemVerilog \textbf{Lines of Code (LOC)} as a proxy metric for design volume. This metric effectively reflects the logical complexity, hierarchical depth, and the scale of interfaces, state machines, and data paths within each module. While acknowledging that coding styles may introduce minor variances, LOC remains a reliable indicator for relative complexity comparisons across diverse IP cores.

\subsection{Quantitative Scale and Distribution}
Our dataset incorporates \textbf{37 distinct IPs} from the OpenTitan repository, totaling approximately \textbf{237,796 lines} of SystemVerilog code. The scale of individual IPs varies significantly, ranging from large-scale modules such as \texttt{spi\_device} (\textbf{32,155 LOC}) to smaller-scale adaptation and glue logic (hundreds to thousands of LOC), mirroring the tiered organizational structure typical of industrial System-on-Chip (SoC) designs.

Statistically, the dataset exhibits a \textbf{median} IP size of 4,482 LOC with an \textbf{interquartile range (IQR)} of 1,906--8,954 LOC. 
As summarized in Table~\ref{tab:opentitan_complexity_dist}, the distribution spans several complexity tiers, where 16 IPs (51.6\%) exceed 5,000 LOC and 8 IPs (25.8\%) surpass the 10,000 LOC threshold. 
This tiered structure reflects the ``long-tail'' distribution prevalent in production-level SoCs, which typically consist of a few high-complexity core modules supported by a larger population of mid-to-small-scale peripheral and control units.

\subsection{Functional Diversity}
Beyond raw scale, the dataset provides comprehensive coverage of critical SoC subsystems, ensuring functional diversity and industrial representativeness. Table~\ref{tab:functional_diversity} categorizes the included IPs into five key architectural domains, spanning from security primitives to complex system infrastructure.

\begin{table*}[tbp]
\centering
\small
\begin{tabularx}{\textwidth}{@{}l l X@{}}
\toprule
\textbf{Subsystem Category} & \textbf{IP Modules} & \textbf{Functional Description} \\ \midrule
\textbf{Cryptography} & \texttt{aes}, \texttt{hmac}, \texttt{kmac}, & Symmetric encryption, hashing, key management, \\
\& \textbf{Security}  & \texttt{keymgr}, \texttt{csrng} & and hardware-based entropy sourcing. \\ \addlinespace[0.5em]

\textbf{Computing}    & \texttt{otbn} & High-complexity Big Number accelerator for public-key cryptography. \\ \addlinespace[0.5em]

\textbf{Infrastructure} & \texttt{tlul}, \texttt{prim} & On-chip bus protocols (TileLink) and fundamental hardware primitives. \\ \addlinespace[0.5em]

\textbf{Storage \&}   & \texttt{flash\_ctrl}, \texttt{otp\_ctrl}, & Controller logic for non-volatile memory, SRAM, \\
\textbf{Lifecycle}    & \texttt{lc\_ctrl}, \texttt{sram\_ctrl} & and hardware lifecycle state management. \\ \addlinespace[0.5em]

\textbf{Peripherals \&} & \texttt{spi\_host/device}, \texttt{i2c}, & Standard communication protocols and SoC-level \\
\textbf{Debugging}    & \texttt{uart}, \texttt{usbdev}, \texttt{rv\_dm} & debug/trace infrastructure (JTAG/RISC-V DBG). \\ \bottomrule
\end{tabularx}
\caption{Functional Diversity and Subsystem Representation}
\label{tab:functional_diversity}
\end{table*}

The combination of significant code volume, varied complexity gradients, and multi-domain functional coverage ensures that our dataset is representative of authentic industrial engineering environments. This provides a robust foundation for evaluating and validating methods under realistic hardware design constraints.

\section{EvoRTL-Bench}
\label{app:EvoRTL}
\subsection{Prompt for \texorpdfstring{$\Delta$Spec}{DeltaSpec} Extraction}
\label{app:prompt-deltaspec}


To transform free-form commit messages into structured change specifications, we prompt an LLM to produce a normalized $\Delta$Spec in a single JSON object. 
The input is the raw Git commit message (treated as a natural-language change request). 
The output schema contains: (i) a coarse-grained change label (\texttt{Functional}/\texttt{Non-Functional}/\texttt{Unclear}), (ii) a confidence score reflecting behavioral clarity and internal consistency, (iii) a brief rationale, and (iv) a structured behavioral description including prior/revised summaries ($S_{\text{old}}$, $S_{\text{new}}$), functional \texttt{Context}, and design \texttt{Intention}.

We explicitly instruct the model to avoid inferring functionality from module/signal names or domain priors: \texttt{Context} must only restate behavior that is explicitly mentioned in the commit message, and should be left empty otherwise. 
Similarly, $S_{\text{old}}$/$S_{\text{new}}$ are constrained to be concise (at most two sentences each) and must not include file paths or line numbers. 
These constraints are designed to reduce hallucination and improve reproducibility across commits with heterogeneous writing quality.

Listing~\ref{lst:system-prompt} shows the system prompt used for $\Delta$Spec extraction (the commit message is provided as the user input).

\begin{lstlisting}[basicstyle=\ttfamily\footnotesize,breaklines=true,caption={System prompt for $\Delta$Spec extraction.},label={lst:system-prompt}]
You are an expert hardware specification analysis assistant.
Your input is a Git commit message describing RTL or hardware design changes.
Your task is to analyze it and return a structured JSON object.

Classification criteria:
- "Functional": The change affects externally observable or software-visible behavior
  (e.g., new modes, states, interrupts, alerts, thresholds, bit-widths, register fields,
  handshake protocols, clock/reset/idle/timing semantics, or architectural functionality).
- "Non-Functional": The change is purely refactoring, code movement, renaming, formatting,
  CI/scripts, comments, documentation, or testbench-only fixes.
- "Unclear": The message is too short or lacks sufficient behavioral detail
  to generate a reliable before/after summary.

Confidence score (0-1):
- Behavioral Clarity (0-0.8): How clearly the message describes what changed,
  what module or signal it affects, and under what condition.
- Consistency (0-0.2): Whether the message is internally consistent and unambiguous.
confidence = Clarity + Consistency.

Your tasks are:
1. Classify the message as "Functional", "Non-Functional", or "Unclear".
2. Provide a "confidence" score from 0 to 1.
3. Provide a concise "rationale" for the classification and score.
4. Generate concise "S_old" and "S_new" summaries (max 2 sentences each).
   - They should describe the module behavior before and after the change.
   - Do NOT include file paths or line numbers.
5. Extract "Context"
   - Summarize only the functional points or behavioral aspects of the affected module or its submodules that are explicitly mentioned in the commit message.
   - Do NOT infer or assume any functionality based on module names, signal names, or general domain knowledge. If the commit message does not explicitly describe any internal functionality or behavior, set this field to an empty string ("").
   - Do NOT describe what was changed, added, or fixed; those belong in "S_old" and "S_new".
6. Extract "Intention" - the design motivation or purpose of the change,
   such as fixing timing issues, adding safety logic, improving coverage, or enabling new functionality.

You MUST output only a single valid JSON object in the exact format below:

{
  "label": "",
  "confidence": 0.0,
  "rationale": "",
  "Context": "",
  "Intention": "",
  "S_old": "",
  "S_new": ""
}
\end{lstlisting}

\subsection{Quality Control and Filtering}
\label{sec:qc_appendix}

To ensure EvoRTL-Bench reflects real-world functional evolution while remaining reproducible and low-noise, we applied a multi-stage Quality Control (QC) pipeline combining deterministic rule checks, diff- and AST-based consistency checks, and manual auditing.

\paragraph{QC Phase 1: $\Delta$Spec Extraction.}
Given the raw commit message, the LLM is prompted to output a single JSON object. We enforce:
\begin{enumerate}[nosep, leftmargin=*] 
    \item \textbf{Schema and type validation.} The output must be valid JSON with keys: \texttt{label}, \texttt{confidence}, \texttt{rationale}, \texttt{Context}, \texttt{Intention}, $S_{\text{old}}$, $S_{\text{new}}$. We validate types (e.g., numeric \texttt{confidence} $\in [0,1]$) and restrict \texttt{label} to \{\texttt{Functional}, \texttt{Non-Functional}, \texttt{Unclear}\}. Malformed outputs are discarded if lightweight repairs fail.
    \item \textbf{Format constraints.} $S_{\text{old}}$ and $S_{\text{new}}$ must be concise (max 2 sentences) and devoid of file paths or line numbers (filtered via regex). For \texttt{Functional} samples, summaries cannot be empty.
    \item \textbf{Content separation.} \texttt{Context} must describe \textit{pre-existing} behavior, not the change itself. We apply heuristics to flag edit verbs (e.g., ``add'', ``fix'') in \texttt{Context}; violators are cleared or discarded.
    \item \textbf{Anti-hallucination checks.} We compare entities in \texttt{Context}/\texttt{Intention} against the commit message. Samples introducing many unsupported technical entities (signals/modules) are filtered to prioritize precision over recall.
    \item \textbf{Label-based filtering.} Only \texttt{Functional} commits are retained. \texttt{Non-Functional} (formatting, CI, docs) and \texttt{Unclear} cases are excluded. Low-confidence scores trigger stricter auditing.
\end{enumerate}

\paragraph{QC Phase 2: Diff- and AST-based Block Ground Truth.}
For retained commits, we map line-level diff hunks to enclosing AST scopes to identify affected RTL blocks:
\begin{enumerate}[nosep, leftmargin=*]
    \item \textbf{Source-file eligibility.} Commits must modify at least one SystemVerilog source (`.sv`, `.svh`), excluding testbench-only changes.
    \item \textbf{Successful localization.} Every diff hunk must map to a valid syntactic scope (e.g., `assign`, `always\_ff`). Commits with unparsable or unmapped hunks are discarded.
    \item \textbf{Deduplication.} We normalize block identifiers to produce a unique set of affected blocks per commit, removing overlaps.
    \item \textbf{Sanity checks.} Excessively large affected sets (indicative of massive refactoring) are flagged for manual review or removal.
\end{enumerate}

\paragraph{Manual Spot Auditing.}
We conducted manual spot auditing on a random sample of $N=100$ commits, stratified across IPs and confidence levels. Hardware-literate annotators reviewed samples against diffs/messages. We evaluated: (i) label correctness, (ii) behavioral fidelity of $S_{\text{old}}/S_{\text{new}}$, (iii) grounding of \texttt{Context} (no change narration), (iv) support for \texttt{Intention}, and (v) localization consistency.
Overall, 94 of 100 samples passed all criteria. Common issues included minor hallucinations in \texttt{Context} or excessive detail in summaries; these subsets were re-processed with stricter constraints.
\begin{table}[h]
\small
\centering
\begin{tabular}{l r}
\toprule
Criterion & Pass rate (\%) \\ \midrule
Label correctness                               & 98.0 \\
$S_{\text{old}}/S_{\text{new}}$ behavioral fidelity & 95.0 \\
\texttt{Context} grounded (no change narration) & 96.0 \\
\texttt{Intention} supported by message         & 99.0 \\
Block localization consistent with diff/AST    & 97.0 \\ \bottomrule
\end{tabular}
\caption{Manual audit checklist and outcomes.}
\label{tab:audit_results}
\end{table}

\paragraph{Outcome.}
After QC, each sample contains a verified $\Delta$Spec and deduplicated affected RTL blocks, enabling multi-positive supervision for real-world RTL evolution.

\section{Offline Overhead and DTG Statistics}
\label{app:offline_overhead}

\paragraph{Offline DTG construction overhead.}
Constructing the Design-Wide Topology Graph (DTG) is a one-time offline step performed for each RTL snapshot prior to query-time
localization, and the resulting graphs are cached and reused across many change requests.
We measure DTG construction time over 25 representative snapshots from EvoRTL-Bench on a standard workstation.
The overhead is modest, ranging from 0 s to 110 s, with a \textbf{median of 9 s} and an \textbf{average of 17.1 s}, indicating that
DTG construction can be integrated into typical development workflows and amortized over repeated queries.

\begin{table}[t]
\centering
\small
\setlength{\tabcolsep}{4pt}
\renewcommand{\arraystretch}{1.1}
\begin{tabular}{l c}
\toprule
\textbf{Statistic} & \textbf{Time (s)} \\
\midrule
\#Snapshots & 25 \\
Min / Median / Mean & 0 / 9 / 17.1 \\
P90 / P95 & 33.2 / 54.8 \\
Max & 110 \\
Total & 428 \\
\bottomrule
\end{tabular}
\caption{Offline DTG construction time over 25 representative snapshots.}
\label{tab:dtg_build_time}
\end{table}

\paragraph{DTG scale diversity.}
Table~\ref{tab:dtg_stats_summary} summarizes DTG scale across OpenTitan IPs.
For each IP, we report the number of extracted DTG instances (\textit{Snapshots per IP}) and the average DTG size
(\textit{Avg Nodes/Edges per IP}, computed over snapshots within that IP).
DTG sizes vary substantially with design complexity, spanning from lightweight control logic to large compute-intensive IPs.
The distribution is right-skewed: most IPs exhibit moderate graph sizes, while a few complex IPs (e.g., \texttt{otbn}, \texttt{kmac})
contribute to a much larger mean edge count.

\begin{table}[t]
\centering
\small
\renewcommand{\arraystretch}{1.15}
\setlength{\tabcolsep}{4pt}
\begin{tabularx}{\columnwidth}{@{}Xrrrr@{}}
\toprule
\textbf{Metric} & \textbf{Min} & \textbf{Median} & \textbf{Mean} & \textbf{Max} \\
\midrule
Snapshots per IP & 5 & 71 & 88.0 & 389 \\
Avg Nodes per IP & 38.2 & 278.8 & 418.5 & 1,980.1 \\
Avg Edges per IP & 231 & 18,607 & 173,633 & 1,682,282 \\
\bottomrule
\end{tabularx}
\caption{Statistical summary of DTGs across OpenTitan IPs.}
\label{tab:dtg_stats_summary}
\end{table}